# Exploring Artistic Visualization of Physiological Signals for Mindfulness and Relaxation: A Pilot Study

Zihan Xu and Youngjun Cho*

University College London

Abstract: Mindfulness and relaxation techniques for mental health are increasingly being explored in the human-computer interaction community. Physiological signals and their visualization have often been exploited together in a form of biofeedback with other intervention methods. Here, we aim to contribute to the body of existing work on biofeedback interfaces for mindfulness, with a particular focus on incorporating artistic effects into physiological signal visualization. With an implemented artistic biofeedback interface, we conduct a pilot study where 10 participants attend stress-induction sessions followed by two biofeedback mindfulness sessions: classic biofeedback and artistic visualization. The result demonstrates that artistic visualization-driven biofeedback significantly improves the effectiveness of biofeedback in helping users feel relaxed in comparison with a classic graphical form of biofeedback. Also, it shows that the artistic effect makes it easy to understand what biofeedback represents. Future work includes exploring how advanced physiological computing methods can improve its efficiency and performance.

## 1 INTRODUCTION

Mental health disorders are recognized as an urgent global health challenge, affecting 10.7% of the population worldwide (Dattani et al. 2021). Despite this prevalence, 70% of people with mental disorders lack access to well-being services (Henderson et al. 2013). The barriers to accessing treatment services include geopolitical resources, insufficient treatment professionals, financial support and stigma (Weiss and Pollack 2017). Assistive technology for mental health can possibly address these issues. In particular, digital mindfulness tools that use visual interactive elements to encourage self-awareness and self-reflection are considered as accessible and effective tools (Patibanda et al. 2017). Various studies have shown that art can have a direct or indirect stimulating effect on brain activity (Altay et al. 2017). When used in a therapeutic context, art allows unconscious feelings to be expressed, identified, and reflected upon to connect and respond to physical symptoms and feelings to increase emotion regulation and self-awareness (Blomdahl et al. 2016). In addition, the use of biofeedback could further enhance self-perception and self-regulation for mental wellbeing as well as empower empathy and compassion in social biofeedback interaction (Moge et al. 2022).

The use of assistive technology could improve the accessibility, participation, effectiveness, and affordability of mental health treatment (Doherty et al. 2008); however, many existing intervention techniques remain limited due to intrinsic factors involving difficulties with being aware of our own internal states (Botella et al. 2012). In this paper, we explore how artistic visualization of physiological signals can improve self-awareness of our body and effectiveness of biofeedback intervention. Our artistic visualization biofeedback interface builds on the literature on biofeedback, mindfulness and arts for wellbeing. For the rest of this paper, we describe the design process and implementation, and discusses the limitations for future work.

## 2 RELATED WORK

### 2.1 Biofeedback for well-being in HCI

Research has seen a growing use of physiological computing methods to understand mental health needs, as they can be indicative of our bodily functioning and mind associated with the autonomic nervous system activities (Cho et al. 2019; Moge et al. 2022). Visualization of real-time physiological data in a form of biofeedback can be used to enhance individuals' physiological state recognition and guide emotion regulation. Research on biofeedback visualization has focused on how to implement more understandable, attractive, and reliable daily utilization practices (Moge et al. 2022).

Biofeedback visualization encourages awareness and engagement with emotions through creative mediums to facilitate novel perspective-taking. These biofeedback displays can also help people become aware of the physiological effects of mental states and can be used for reflection because they provide a novel narrative that can be integrated into their lived experiences. For instance, *Heart Calligraphy* (Yu et al. 2016) emphasizes the individual uniqueness of HRV patterns by using them to draw abstract portraits of participants, while *Metaphone* (Simbelis and Höök 2013) transforms emotional arousal derived from pulse and skin conductance into colorful paintings.

Interestingly, artistic effects can provide users with an interactive opportunity to engage with changes in their autonomic nervous system when generated dynamically. *Ethereal Phenomena* (Turbay et al. 2022) offers a meditative experience when a symbolic Tibetan artwork becomes an extension of the participant's body when it reacts to their breathing. Other work aims to support participants in exploring connections between different bodily functions and emotion. For instance, *Mettamatics* (Khut and Howard 2020) used dynamic power spectrum density displays of heartbeat data accompanied by sounds that rose and fell according to breathing to show how breathing patterns can affect HRV. In addition to self-oriented socio-emotional skills such as self-awareness and self-reflection, sharing art-based biofeedback has the potential to cultivate interpersonal skills, such as empathic engagement, when shared with others (Moge et al. 2022). In *BioFlockVR* (Song et al. 2019) pairs of participants were immersed in nature visuals that changed according to their EEG, heart rate, and gestures.

**2.2 Mindfulness and Art Therapy for well-being in HCI**

Studies have suggested that mindfulness-based breathing exercises can improve emotion regulation by guiding people to concentrate on their bodily sensations (Patibanda et al. 2017; Prpa et al. 2018). Further, a recent study has demonstrated that digital personalized mindfulness tools are more beneficial for self-awareness and their use is increasing as apps use embedded sensors to guide meditation [9].

Art itself has also been proposed as an alternative treatment for mental disorders. Research has investigated the role of artistic expression to help people reflect on positive and negative emotions (Greenberg and Watson 2006; Wagener et al. 2022). Art therapy is both a physical expression of emotional experiences and a record of the dynamic journey of well-being (Havsteen-Franklin et al. 2021). These can be achieved by engaging in image-making activities rather than conversation because internal dialogue begins to form a space for self-reflection (Blomdahl 2017). The shapes and colors of different patterns can be used to express the nuanced personalities of the patients. Regarding art therapy design, the Expressive Therapy Continuum (ETC) and the Media Dimension Variable can be considered as guidance, with non-verbal communication and creative expression being at the core of art therapy (Ching-Teng et al. 2019). Based on ETC, using different sensory mediums can emphasize the transformation of feelings into concrete imagery to release repressed emotions.

Building upon the literature, we aim to answer our research question: *Does artistic visualization influence the effectiveness of biofeedback in mindfulness intervention and relaxation?* We attempt to answer this with the design of a novel interface for art-based biofeedback. We describe the design process, the details of our experimental protocol, and findings. We conclude by discussing limitations and future work.

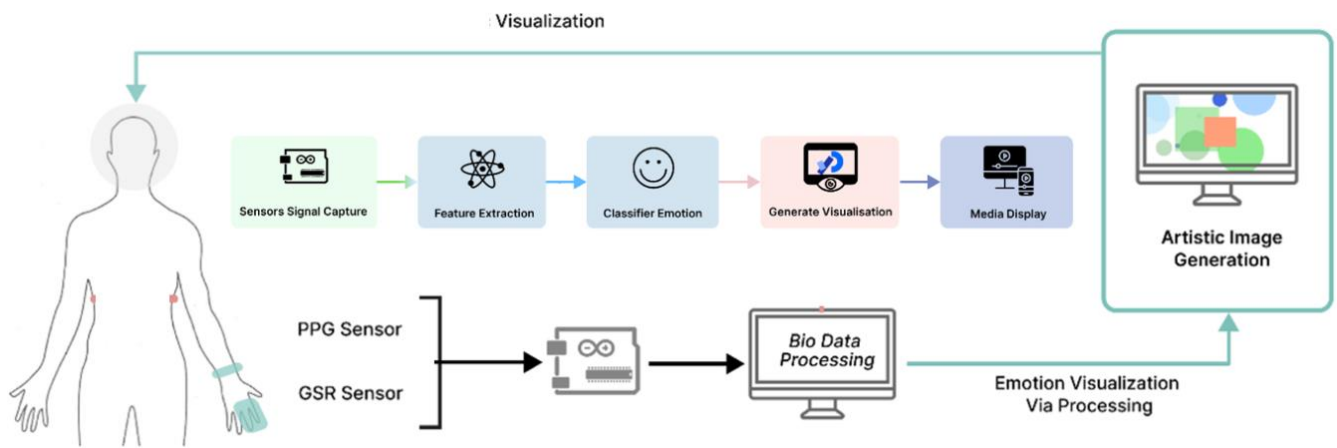

Figure 1. Artistic visualization Biofeedback loop design and steps to render feedback

## 3 PROPOSED ARTISTIC VISUALIZATION BIOFEEDBACK

Figure 1 details our proposed artistic visualization-driven biofeedback interface that maps physiological data to specific emotions and generates them into artistic visualizations. The prototype consists of three steps including physiological data collection, data processing, and data visualization.

### 3.1 Measuring and classifying emotions

Changes in skin conductance and heart rates are indicative of the autonomic nervous system activities. We mainly use galvanic skin response (GSR) and photoplethysmography (PPG) sensors (see Figure 1) which have often been used to infer emotional states (Udovičić et al. 2017). The collected data from GSR and PPG sensor are processed into three categories of emotional states according to the three classes of valence and arousal: positive, neutral, and negative with a shallow machine learning classifier (Ayata et al. 2017). Based on these classes of valence and arousal, emotions can be further classified into happiness, peacefulness, sadness, and fear (Küller et al. 2009).

### 3.2 Artistic Visualization of physiological signatures of emotion

With an opensource physiological computing toolkit (Joshi et al. 2023), we generate visual presentations (as shown in Figure 2) of emotions according to Kandinsky's principle, which proposes that sharp forms (e.g. triangle) and round form (e.g. circle) can emphasize different types of feelings (Dreksler 2020). In addition, certain colors can foster positive emotions (e.g., green) green while others may elicit stress (e.g. red) (Küller et al. 2009).

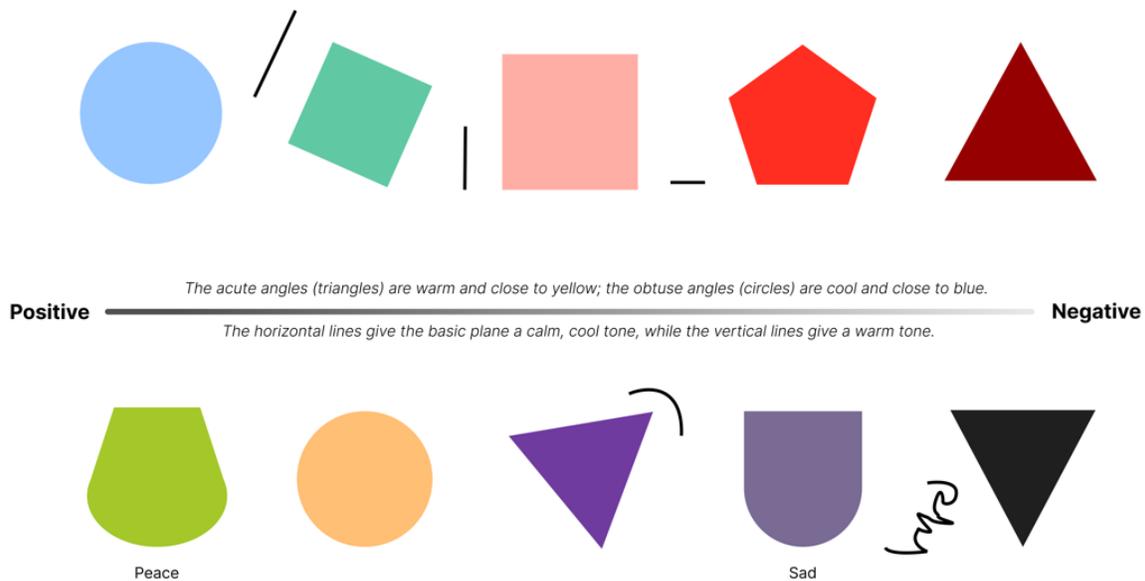

Figure 2. Emotion visualization based on Kandinsky's principle

## 4 STUDY

In this section, we describe our pilot experiment we conducted to investigate the influence of different biofeedback forms for emotion regulation.

### 4.1 Participants and Procedures

To evaluate the usability of the prototype, we conducted our study with 10 volunteer participants (6 females and 4 males, 22-27 years). Each participant received an information sheet and was asked to review and sign a consent form at the start of the experiment. The participants were debriefed that below procedure was approved by the University College London Interaction Centre ethics committee (ID Number: UCLIC/1920/006).

Figure 3 illustrates the experimental process, which lasted about 45 minutes for each participant. Participants started by completing questions about perceived mental stress and affect (i.e., the PANAS scales; Watson et al. 1988). Each participant encountered a stress-inducing task (i.e., the Canabalt game[1]) followed by a counterbalanced emotion regulation condition. There were two emotion regulation conditions in this study to examine the effects of biofeedback on stress: a raw data visualization and our artistic visualization biofeedback interface. The artistic visualization effects were designed to change its patterns along with recognized affects; mainly positive, neutral and negative affects through the physiological computing toolkit (as shown in Figure 4). Participants were asked to wear physiological sensors (GSR, PPG, and respiration) during both emotion regulation tasks. Each task was followed by a set of questions on perceived mental stress and affect (i.e., PANAS). After completing the experimental tasks, participants were invited to fill out a questionnaire to evaluate the usability of the emotion regulation interfaces. The study then was followed by a semi-structured interview during which they were asked about their opinions on the accuracy of the biofeedback, preference for two biofeedback methods, as well as views on the mindfulness based on breathing practice.

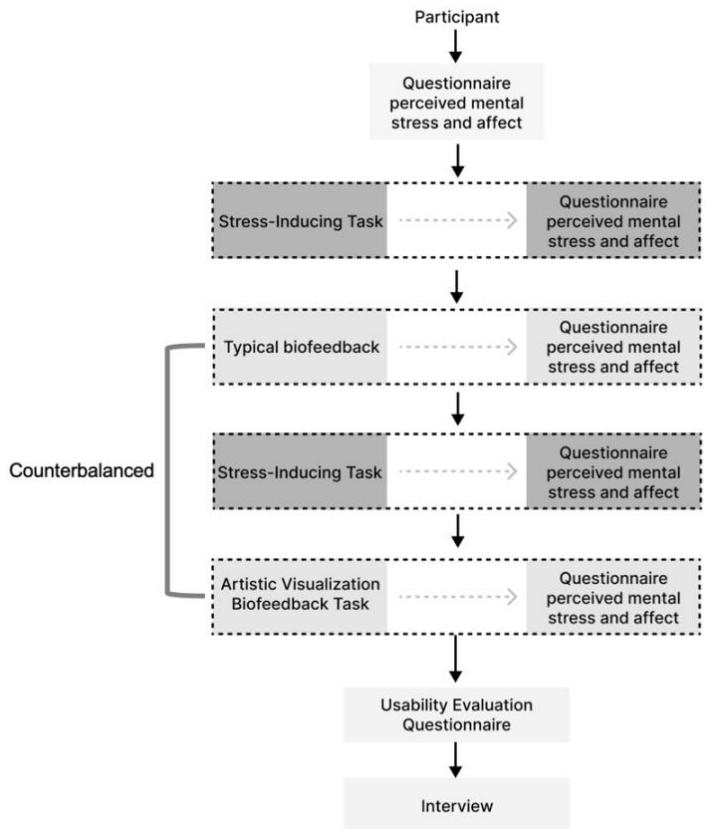
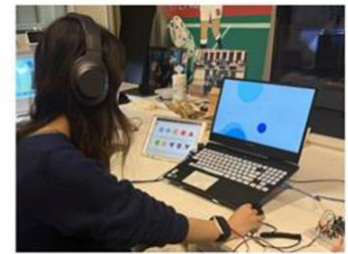
Artistic visualization for Positive State

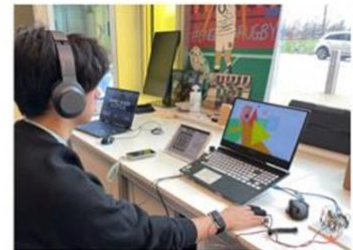
Artistic visualization for Natural State

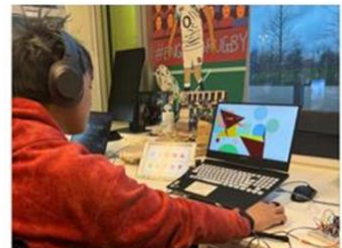
Artistic visualization for Negative State

Figure 3. Experimental protocol     Figure 4. Artistic visualization setups for different states

## 5  RESULTS

### 5.1 Quantitative Analysis

Figure 5 shows the distributions of the self-reported values of the 7 dependent variables (complexity, easy-to-understand, difficulty, relaxation level, willingness to use, Positive affect, Negative affect) for the two conditions (Classic visual biofeedback and artistic visualization biofeedback).

---
[1] https://canabalt.com/

For significance tests, we carried out a paired-samples t-test on complexity and willingness. The results showed a significant effect of the type of interface on willingness (t(9)=-4.071, p=0.003) and approaching significance on complexity (t(9)=2.212, p=0.054). As the data from the remaining variables were not found normally distributed (Shapiro-Wilk test; p<0.05 for all cases), we used the Wilcoxon signed rank test. The test confirmed there was a significant effect on relax (Z=-2.669, p=0.008), indicating that participants felt much more relaxed during the art-based biofeedback. For the variables easy-to-understand, difficulty, there was no significant effect. Lastly, we conducted a paired-samples t-test for the remaining two DVs on affects: positive affect, negative affect. The results showed no significant effect of the biofeedback type on affects (p=0.301 for positive affect, p=0.690 for negative affect) while participants found the art-based biofeedback interface more positive and less negative in general.

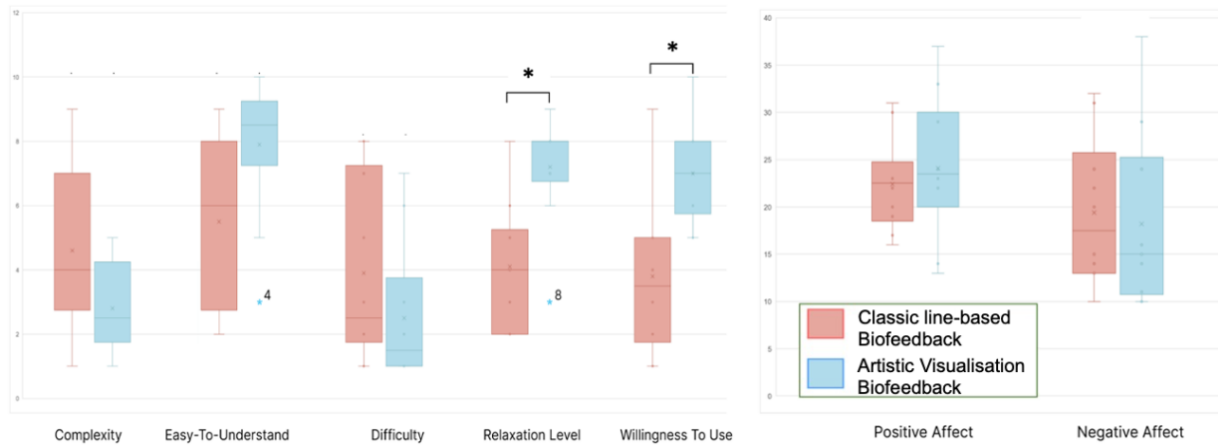

Figure 5. Results of the self-reported values of 7 dependent variables (complexity, easy-to-understand, difficulty, relaxation level, willingness to use, and the PANAS positive and negative affects) for the two conditions of classic visual biofeedback and artistic visualization biofeedback. Statistically significant results are marked with asterisks (*).

### 5.2 Qualitative Analysis

In the interview, almost all participants indicated a preference for the artistic visualization to practice emotion regulation over the typical graphical form of biofeedback (that is made of simple line plots) because the latter can be confusing or complex. Participants found the artistic visualization-driven biofeedback more intuitive and easier to understand and interpret, in addition to being a pleasant and comfortable visual experience. In contrast, participants reported that the rapidly fluctuating lines from the classic visual biofeedback often created a sense of stress.

All participants (N=10) responded that the artistic visualization seemed to accurately reflect their inner emotional states, and it was easy to associate different shapes and colors with relevant emotions. For instance, the contrasts between bright red and cool blue, obtuse circles and acute triangles, all provided direct sensory stimulation to create emotional connections to positive or negative feelings. The colorful, dynamically changing interface also provided users with aesthetic experiences, drawing users' attention to the process of change in body states.

Further, three participants highlighted that the artistic visualization influenced self-awareness. During the experiment, the elevated stress levels of one participant was detected both by our art-based biofeedback interface and her own smartwatch (with stress level monitoring). During the artistic biofeedback, the presence of the dark red triangle representing negative emotions reinforced her feelings of stress and led her to report in a more negative way.

## 6 DISCUSSION & FUTURE WORK

In this paper, we have explored how artistic visualization influences the effectiveness of biofeedback in mindfulness intervention and relaxation. With an opensource physiological computing toolkit (Joshi et al. 2023), we have developed a cost-effective interface to practice emotion regulation though art-based biofeedback. The experimental results have demonstrated that the integration of physiological computing and positive art-therapy can be effective in improving mental wellbeing.

As usual, some limitations of this work should be noted. First, wearing multiple sensors during the process might have affected users' experience, attention, and movement. A possible flexible and accessible solution could involve integrating sensors into a wearable, such as wristband or ring-based interfaces (Kinnunen et al. 2020; Cho et al. 2017). The visual and audio parts could also be displayed in immersive virtual reality to help reduce external distractions and improve attention (Prpa et al. 2018; Wang et al. 2020, 2022), enabling participants to focus more on mindfulness relaxation. Further, the present study focuses only on intrapersonal states, limiting empathic advantages offered by social interactions. Social art-based biofeedback could therefore be introduced to encourage people to create collaboratively with others (Moge et al. 2022). In future work, it would be interesting to investigate how artistic visualization biofeedback could be practiced together with existing effective interventions such as mindfulness-based breathing exercises (Patibanda et al. 2017; Prpa et al. 2018). Also, other types of modality (e.g. haptic; Schmitz et al. 2020) and physiological sensing channels such as respiration (Cho et al. 2017a, 2017b) could be explored to enhance positive attitudes towards biofeedback intervention when being in a negative mood.